\documentclass[journal]{IEEEtran}

\pdfoutput=1

\usepackage{amsmath,bm,amssymb,amsfonts,amsthm}
\usepackage[linesnumbered, ruled, vlined]{algorithm2e}
\usepackage{graphicx}
\usepackage{xcolor} 
\usepackage{relsize}
\usepackage{textcomp}
\usepackage{listings}
\usepackage{gensymb}

\usepackage{changepage}
\usepackage{hhline}
\usepackage{multirow}
\usepackage{nomencl}
\usepackage{mathtools}
\usepackage{commath}
\usepackage{booktabs}
\usepackage{subfiles}
\usepackage{tabularx}
\usepackage{lscape}
\usepackage{rotating}

\usepackage{verbatim}
\usepackage{comment}

\usepackage[normalem]{ulem}
\usepackage[utf8]{inputenc}
\usepackage[hyphens]{url}
\usepackage{longtable}
\usepackage{lineno}
    \modulolinenumbers[5]
\usepackage[caption=false]{subfig}
\usepackage{wrapfig}
\usepackage{enumitem}
    \setlist{nolistsep}
\usepackage[hidelinks]{hyperref}
\usepackage{makecell}
\usepackage{multicol}
\usepackage{colortbl}

\usepackage{scalerel}
\usepackage{tikz}
\usetikzlibrary{svg.path}
\definecolor{orcidlogocol}{HTML}{A6CE39}
\tikzset{
  orcidlogo/.pic={
    \fill[orcidlogocol] svg{M256,128c0,70.7-57.3,128-128,128C57.3,256,0,198.7,0,128C0,57.3,57.3,0,128,0C198.7,0,256,57.3,256,128z};
    \fill[white] svg{M86.3,186.2H70.9V79.1h15.4v48.4V186.2z}
                 svg{M108.9,79.1h41.6c39.6,0,57,28.3,57,53.6c0,27.5-21.5,53.6-56.8,53.6h-41.8V79.1z M124.3,172.4h24.5c34.9,0,42.9-26.5,42.9-39.7c0-21.5-13.7-39.7-43.7-39.7h-23.7V172.4z}
                 svg{M88.7,56.8c0,5.5-4.5,10.1-10.1,10.1c-5.6,0-10.1-4.6-10.1-10.1c0-5.6,4.5-10.1,10.1-10.1C84.2,46.7,88.7,51.3,88.7,56.8z};
  }
}
\newcommand\orcidicon[1]{\href{https://orcid.org/#1}{\mbox{\scalerel*{
\begin{tikzpicture}[yscale=-1,transform shape]
\pic{orcidlogo};
\end{tikzpicture}
}{|}}}}

\usepackage[noadjust]{cite}

\begin{document}

\title{\huge Real-Time Implementation of Dynamic State Estimation \\ for Microgrid Load Bus Protection}

\author{
    Sarbajit~Basu $^{1}$\orcidicon{0000-0002-2581-3089},
    Arthur~K.~Barnes $^{2}$\orcidicon{0000-0001-9718-3197},
    Adam~Mate $^{2}$\orcidicon{0000-0002-5628-6509},
    and Olga~Lavrova $^{1}$\orcidicon{0000-0000-0000-0000}
    \vspace{-0.25in}

\thanks{Manuscript submitted:~Mar.~3,~2023.
Current version: Apr.~21,~2023.
U.S. Government work not protected by U.S. copyright.
}

\thanks{$^{1}$ The authors are with the Klipsch School of Electrical and Computer Engineering at New Mexico State University, Las Cruces, NM 88003 USA. Email: sarbasu@nmsu.edu, olavrova@nmsu.edu.}

\thanks{$^{2}$ The authors are with the Analytics, Intelligence, and Technology Division at Los Alamos National Laboratory, Los Alamos, NM 87545 USA. \\ Email: abarnes@lanl.gov, amate@lanl.gov.}

\thanks{LA-UR-22-29464. Approved for public release; distribution is unlimited.}

}

\markboth{2023 IEEE Kansas Power and Energy Conference, April~2023}{}

\maketitle


\begin{abstract}
Inverter-interfaced microgrids, owing to the lack of fault current, cannot be protected using traditional over-current protections, while admittance or differential relaying protection schemes are not practical to be implemented.
Dynamic state estimation can track and predict power system transients and has been extensively investigated for setting-less protection.
A novel real-time application of dynamic state estimation for protection is proposed in this paper, wherein parameter estimation and parallel processing is used to identify the state of the system. The implementation scheme has low process complexity and employs a data acquisition device and estimator that run on a general-purpose computer.
This proposed implementation extends the state-of-the-art, under short-circuit conditions, to a real-time implementation with a lumped-load radial microgrid and a grid-forming inverter with current-limiting behavior.
\end{abstract}


\section{Introduction} \label{sec:introduction}
\indent

Owing to their smaller carbon footprint, renewable resource-based generation -- often inverter-interfaced -- is becoming increasingly prevalent. However, due to their rapid current-limiting behavior, these resources requires novel protection schemes.
Traditional protection schemes are unable to detect faults on microgrids, require complex coordination schemes among relays, and are susceptible to misoperation owing to erroneous values provided by faulty measurement devices \cite{Liu2021Dynamic, Telukunta2017ProtectionCU}. As inverter-dominated microgrids are unable to supply high currents during faulted conditions, the inverters shut down during faults. Additionally, bi-directional flow of current during switching over from grid-connected to islanded mode can make fault detection difficult \cite{Barnes2021Optimization, Dehghanpour2022Protection}. 

Dynamic state estimation (DSE), coupled with high accuracy GPS-synchronized measurements, can accurately track system dynamics and provide details about system operating characteristics. The robustness, combined with the relatively low time complexity of DSE, has paved the way to explore its real-time applications for control and protection:
DSE has been explored for real-time parameter identification to develop dynamic load models and load model behaviors \cite{Liu2021Dynamic, liu2009dynamic, Lin1993Dynamic, Rouhani2016Dynamic}; furthermore,
DSE has been implemented as setting-less protection scheme, using real-time measurement data, for the protection of microgrids \cite{Choi2017Effective, Choi2013Autonomous}.

DSE has also been explored for other setting-less protection applications on both transmission and distribution networks \cite{Barnes2022Inverter, Xie2019Dynamic, Meliopoulos2017Promise, Zhao2019Definition}; it promises to eliminate or mitigate the protection issues described earlier.
Owing to these advantages, DSE performs better in protection of transmission lines, transformers, and capacitor banks compared to distance relaying or other traditional protection methods \cite{Liu2015Microgrid, Xie2020Reliable}.
Finally, DSE-based protection schemes, with access to only the voltage and current measurements on terminals of protected devices, has been investigated for the protection of microgrids while identifying hidden failures in substations \cite{Liu2020HighPenetration, Albinali2017Hidden}.

The complexities of real-time state estimation, however, are several, including but not limited to the availability of secure and reliable channels \cite{Kurt2020Secure}, the volume of data to be considered for estimation \cite{Lin2023Design}, and the pre-processing, handling, and storage of data received from the measurement devices \cite{Chai2015Information}.

This paper introduces a real-time implementation of a state-of-the-art DSE algorithm for adaptive load bus protection in an inverter-interfaced microgrid.
Earlier work focused on ideal voltage sources \cite{Barnes2021Dynamic}, whereas this uses an inverter model with current-limiting behavior \cite{Vasquez2013Modeling} that feeds a balanced passive load.
Most work on DSE considers one load model and distinguishes the normal operational condition from the observed conditions based on the chi-squared confidence; in this work, multiple load models -- corresponding to different operational conditions -- are designed, which clearly designate true operational state, making this implementation more robust.
A developed orchestrator module identifies the relevant operational conditions from estimated system parameters, which in turn are applied to trip the appropriate breakers.
All of these were implemented on a general-purpose computer, and experimental results verify that the proposed method accurately identifies the correct model with low time delay and computational complexity.

\section{Problem Formulation for DSE} \label{sec:discussion}
\indent

\subsection{Grounded-Wye Load with Three-Phase Fault}
\indent

\begin{figure}[!htbp]
\centering
\includegraphics[width=0.4\textwidth]{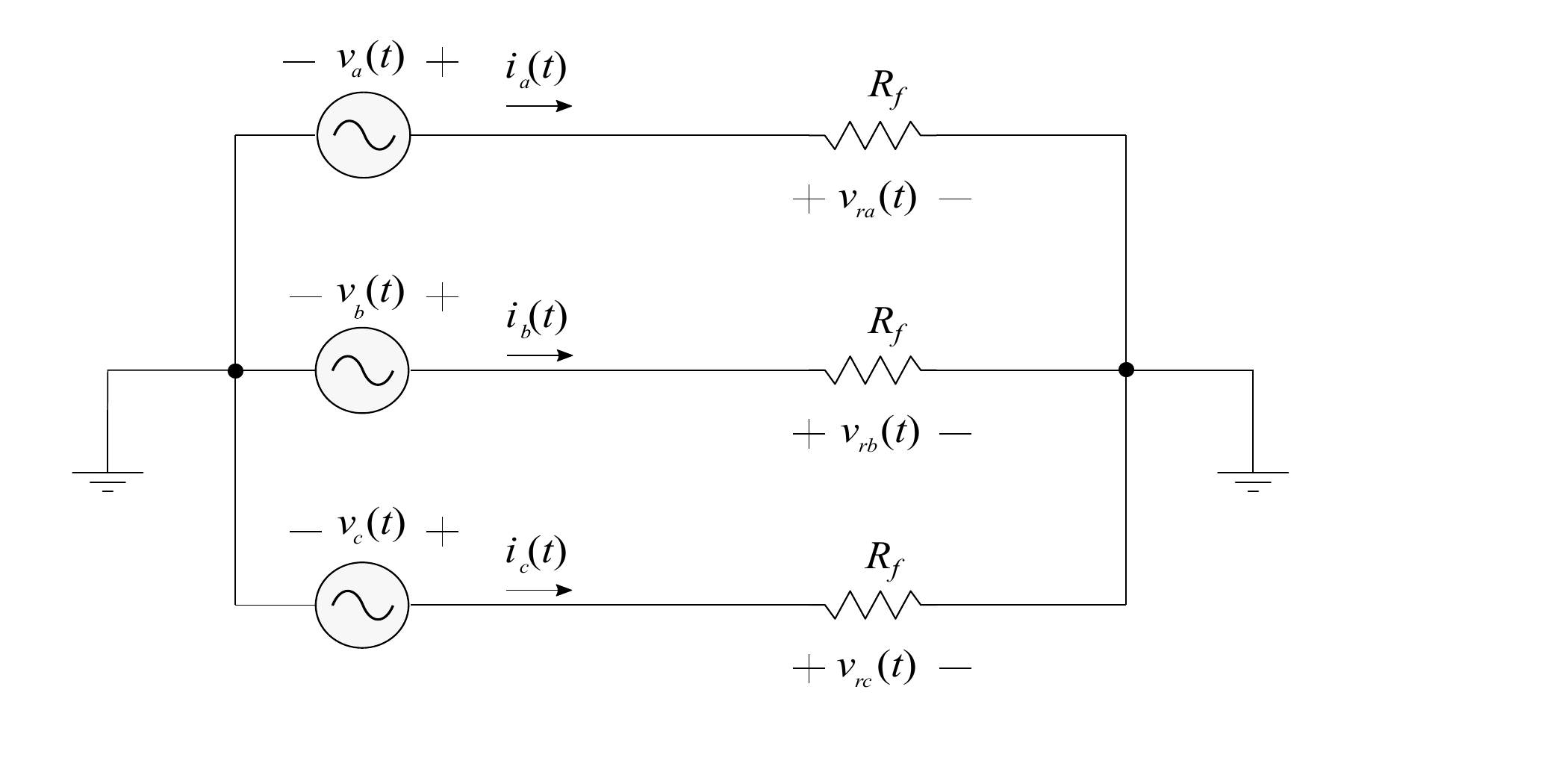}
\caption{Grounded-wye-connected RL load with a three-phase fault.}
\label{fig:3ph-model}
\end{figure}

The problem formulation for the simplified grounded-wye-connected RL load with a three-phase fault, displayed on Fig.~\ref{fig:3ph-model}, is described below.
\textit{For other load configurations, refer to \cite{Barnes2021Dynamic}.}
\vspace{0.1in}

The terminal equations for the case:
\small
\begin{subequations} 
\begin{align}
v_{a}(t) = v_f(t) \quad
v_{b}(t) = v_f(t) \quad
v_{c}(t) = v_f(t) \quad
\end{align}
\begin{align}
i_a(t) = G_f v_{ra}(t) \quad
i_b(t) = G_f v_{rb}(t) \quad
i_c(t) = G_f v_{rc}(t) \quad
\end{align}
\end{subequations}
\vspace{0.05in}

\normalsize
The output and state of the system, respectively:
\small
\begin{subequations}
\begin{align}
y &=    \begin{bmatrix}
        v_a(t) & v_b(t) & v_c(t) & i_a(t) & i_b(t) & i_c(t)
        \end{bmatrix}^T \\
x &=    \begin{bmatrix}
        G_f & v_{ra}(t) & v_{rb}(t) & v_{rc}(t)
        \end{bmatrix}^T
\end{align}
\end{subequations}
\vspace{0.05in}

\normalsize
Assuming that the signals are sampled at timesteps $n \in \{1,2,\ldots,N\}$ with sample time $\Delta t$, the discrete-time state-output mapping function, $h(x)$, is therefore:

\small
$\forall n \in \{1,\ldots,N\}$
\begin{align}
\begin{split}
h_n(x) &= v_{a}(n) = v_{ra}(n) \\
h_{n+N}(x) &= v_{b}(n) = v_{rb}(n) \\
h_{n+2N}(x) &= v_c(n) = v_{rc}(n) \\
h_{n+3N}(x) &= i_a(n) = G_f v_{ra}(n) \\
h_{n+4N}(x) &= i_b(n) = G_f v_{rb}(n) \\
h_{n+5N}(x) &= i_c(n) = G_f v_{rc}(n) \\
\end{split}
\end{align}

\normalsize
The Jacobian of the system, $\mathbf{H}$, is therefore:

\small
$\forall n \in \{1, \ldots, N\}$
\begin{align}
\begin{split}
\frac{\partial v_{a}(n)}{\partial v_{ra}(n)} &= 1 \qquad
\frac{\partial v_b(n)}{\partial v_{rb}(n)} = 1 \qquad
\frac{\partial v_c(n)}{\partial v_{rc}(n)} = 1 \\
\frac{\partial i_a(n)}{\partial G_f} &= v_{ra}(n) \qquad 
\frac{\partial i_b(n)}{\partial v_{rb}(n)} = G_f \qquad
\frac{\partial i_c(n)}{\partial G_f} = v_{rc}(n) \\
\frac{\partial i_a(n)}{\partial v_{ra}(n)} &= G_f \qquad
\frac{\partial i_b(n)}{\partial G_f} = v_{rb}(n) \qquad
\frac{\partial i_c(n)}{\partial v_{rc}(n)} = G_f 
\end{split}
\end{align}

\begin{align*}
\frac{\partial v_{a}(n)}{\partial v_{ra}(n)} &= 1  \qquad
\frac{\partial v_b(n)}{\partial v_{rb}(n)} = 1     \qquad
\frac{\partial v_c(n)}{\partial v_{rc}(n)} = 1  \\
\frac{\partial i_a(n)}{\partial v_{ra}(n)} &= G_f \qquad
\frac{\partial i_b(n)}{\partial v_{rb}(n)} = G_f  \qquad
\frac{\partial i_c(n)}{\partial v_{rc}(n)} = G_f \\
\frac{\partial i_a(n)}{\partial G_f} &= v_{ra}(n) \qquad
\frac{\partial i_b(n)}{\partial  G_f} = v_{rb}(n) \qquad
\frac{\partial i_c(n)}{\partial G_f} = v_{rc}(n) \\
\end{align*}

\normalsize
\textit{Note that for simplicity of notation, the indexing of individual elements of $\mathbf{H}$ is not presented.}

\vspace{0.1in}

\subsection{DSE Solver Formulation}
\indent

This process is repeated iteratively until either the maximum number of iterations is reached or the algorithm has converged, indicated by the change in the log of the squared error falling below a specified threshold:

\small
\begin{equation}
\Delta J_i = \abs{\log \abs{\epsilon_i^* \cdot \epsilon_i} - \log \abs{\epsilon_{i-1}^* \cdot\epsilon_{i-1}}}
\end{equation}
\vspace{0.05in}

\normalsize
The measurement error test is performed as follows:
\small
\begin{equation}
p = F_{m-n}(J_i) \ge 0.95
\end{equation}
\normalsize
\noindent where $F_{m-n}$ is the chi-squared cumulative distribution function for $m - n$ degrees of freedom, in which $m - n$ is the number of linearly independent observables.
\vspace{0.1in}

\section{Real-Time Implementation} \label{sec:methodology}
\indent

The algorithm presented below was implemented as a collection of programs in an interpreted language: separate programs to estimate each load model were developed that run in parallel; these were executed on a general-purpose computer running a POSIX operating system.
The emphasis of the implementation was on developing a system of communication that closely resembles industry standards.

\subsection{Communication of Data}
\indent

The load models were simulated offline and the measurements for different operational conditions were stored in a comma-separated value (CSV) file.
A program was written in the Go language to emulate the behavior of an analog-digital converter (ADC); this program reads data from the CSV files and directs them to the STDIN channel of the orchestrator program, pausing after each line for a time equal to the sample period.
The measurements include time-stamped voltage and current values, with data being sent line by line from the CSV.

\subsection{Orchestrator}
\indent

The orchestrator program receives and re-directs the measurements.
Load models, corresponding to different operational conditions, have been developed in the C language. 
The orchestrator scans all the measurements from STDIN and directs them to the different models, which then executed in parallel as different processes.
The communication is achieved by using two-way channels via Unix pipes between the orchestrator and each model.
Once each model was run, the chi-squared confidence and estimated parameters are being returned to the orchestrator, which then scans all values and selects the model with the highest confidence; after this is completed, the corresponding protective action is recommended and instructions are sent to the breakers to protect the inverter and load.

\begin{figure}[!htbp]
\centering
\includegraphics[width=0.475\textwidth]{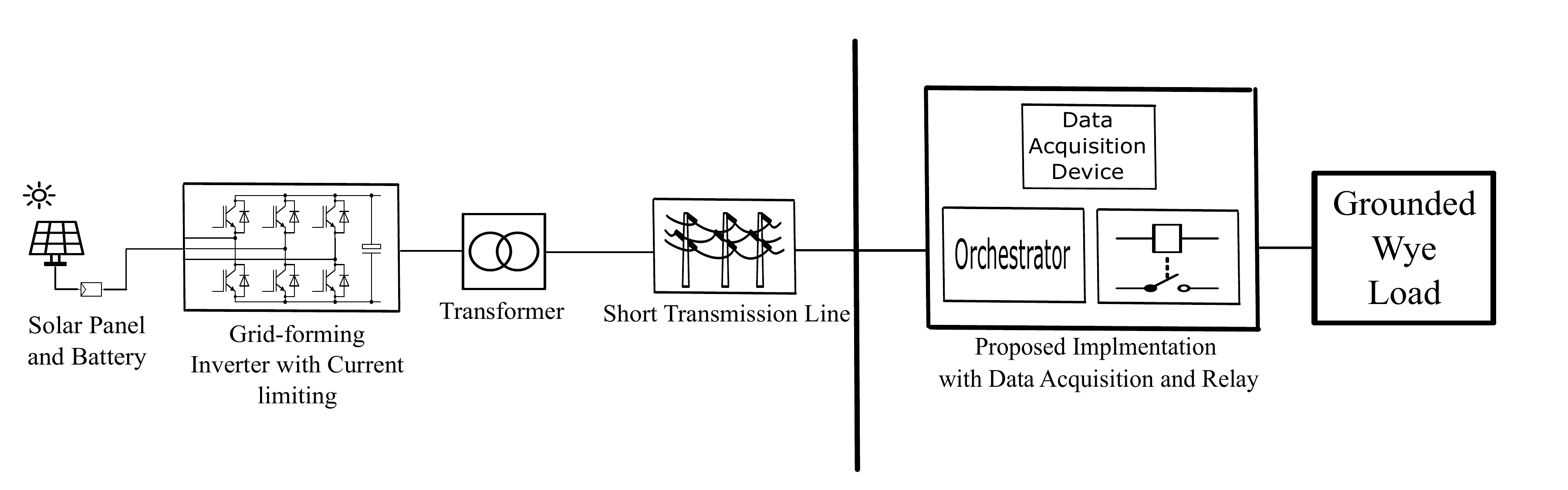}
\caption{Test system used for implementation.}
\label{fig:test-system}
\end{figure}

\small
\begin{algorithm}
\caption{Model Implementation}
\label{alg:model-pseudocode}
\DontPrintSemicolon
    Initialize BUFFER based on window size \;
    \While{reading end of PIPE is non-empty}{
      Read measurements from PIPE \;
      Write measurements to \textit{writing end} of BUFFER \;
         \If{BUFFER is  full} {    
            Run estimator \;      
            Calculate confidence \;
            Write \textit{confidence} to reading end of PIPE \;
           }    
            Remove measurements for oldest \textit{timestamp} from BUFFER\;
    }
\end{algorithm}
\normalsize

\small
\begin{algorithm}
\caption{Orchestrator Implementation}
\label{alg:Orchestrator-pseudocode}
\DontPrintSemicolon
    Spawn subprocesses for each model \;
    Create PIPEs for each model \;
    \While{STDIN is non-empty}{
        Read measurements from STDIN \;
        \For {each model[i]}{  
            Write measurements to \textit{writing end} of PIPE[i] \;        
            Read \textit{confidence} for each model from \textit{reading end} of PIPE[i] \;
        }
        Identify model with highest \textit{confidence} \;
        \If{ new chosen model $=$ old chosen model} {
            $N \leftarrow N + 1$\;
        }
        \Else{
            $N \leftarrow 0$ \;
        }
        \If{ N $>$ \textbf{Hysteresis Samples}} {
            Apply protection settings for model with highest confidence \;
        }
        \Else {
            Apply protection settings for previously chosen model \;
            }
        }
\end{algorithm}
\normalsize

\section{Case Study and Experiments} \label{sec:case-study}
\indent

To verify the implementation of the proposed method and demonstrate the use of DSE in efficiently providing setting-less protection for microgrids, a system with a balanced grounded-wye constant-impedance load was considered; Table~\ref{table1:parameters} summarizes the details of the system.
The simulation was done in MATLAB/Simulink\textsuperscript{\textregistered}, using the Specialized Power Systems library on 64-bit MATLAB R2019b\textsuperscript{\textregistered} \cite{MATLAB:2019}.
The load itself was connected to an inverter rated at 480 $V_{rms}$ line-line, through a 1000~ft of 1/0 AWG quadruplex overhead service drop cable, as demonstrated in Fig.~\ref{fig:test-system} (\textit{on Page~\pageref{fig:test-system}}).

\begin{table}[!htbp]
\centering
\setlength{\tabcolsep}{10pt} 
\renewcommand{\arraystretch}{1.25} 
\caption{Parameters for three-phase dynamic models}
\begin{tabular}{|c|c|} 
\hline
\textbf{Parameter} & \textbf{Values} \\
\hline\hline
Total Load Real Power & 10 kW \\ 
Total Load Reactive Power & 5kVAR \\
Load Resistance (per phase) & 18.432 $\Omega$ \\
Load Inductance (per phase) & 24 mH \\
Fault Resistance & 1 m$\Omega$\\
Ground Resistance & 10 m$\Omega$ \\ 
\hline
\end{tabular}
\label{table1:parameters}
\end{table}

The modes of operation for the three-phase system investigated the following scenarios: 1) Unfaulted; 2) Phase~A line-ground fault; 3) Phase~B line-ground fault; 4) Phase~C line-ground fault; 5) Phases~A-B line-line fault; 6) Phases~B-C line-line fault; 7) Phases~C-A line-line fault; and 8) Three-phase fault.
DSEs were devised corresponding to the modes of operation and run in parallel; the current state of the system was indicated by the DSE with the lowest error.
The fault impedance for each line-ground fault case was 15~m$\Omega$ and the total fault impedance for each line-line fault case was 10~m$\Omega$. For the Three-phase fault, the effective fault impedance was 15~m$\Omega$

\begin{figure}[!htbp]
\centering
\includegraphics[width=0.475\textwidth]{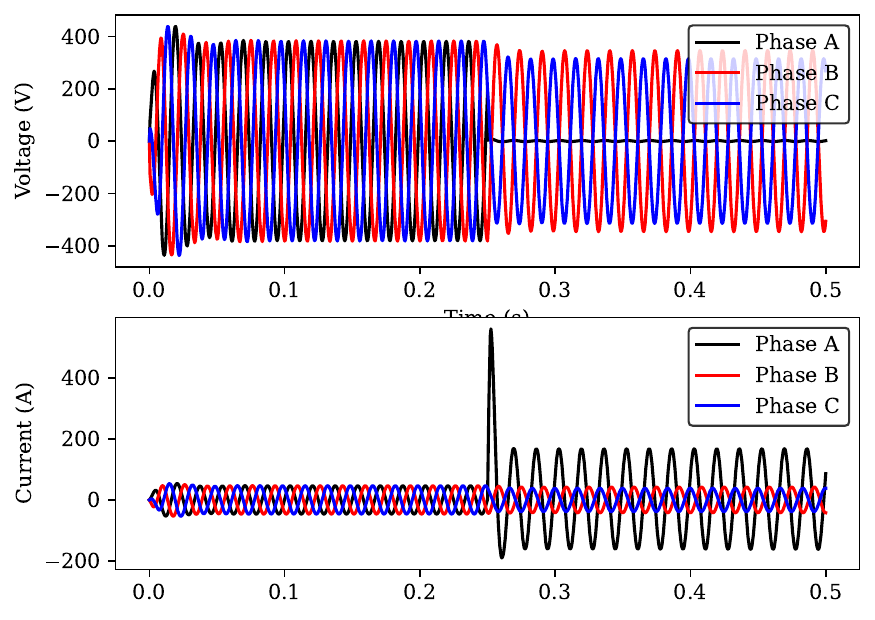}
\caption{Plot of voltage and current for Case A.}
\label{fig:a2g-fault-vi}
\end{figure}

\begin{figure}[!htbp]
\centering
\includegraphics[width=0.475\textwidth]{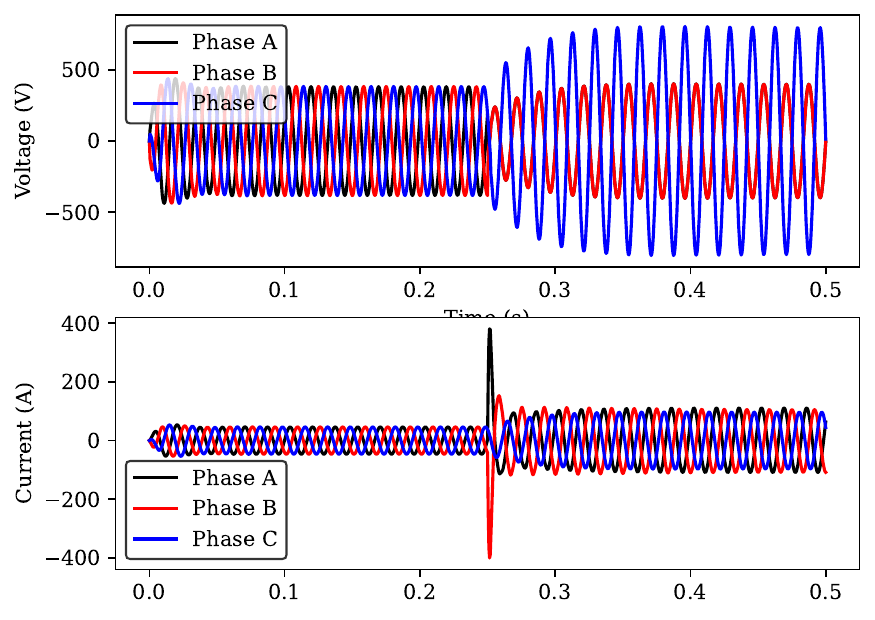}
\caption{Plot of voltage and current for Case B.}
\label{fig:a2b-fault-vi}
\end{figure}

\begin{figure}[!htbp]
\centering
\includegraphics[width=0.475\textwidth]{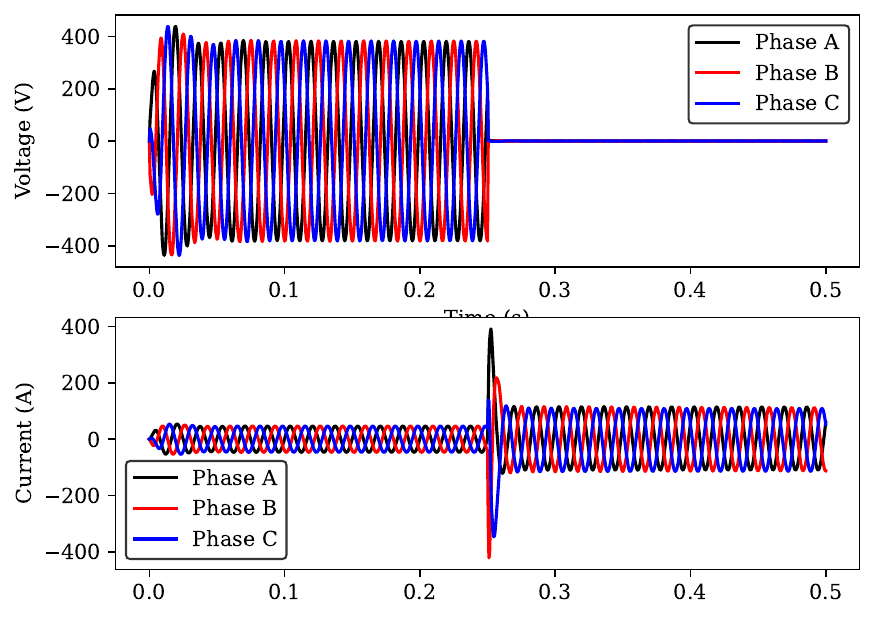}
\caption{Plot of voltage and current for Case C.}
\label{fig:3ph-fault-vi}
\end{figure}

Three different cases (Case~I, Case~II, and Case~III) were devised, as elaborated in Table~\ref{table:Fault-cases}; their corresponding voltage and current variations are demonstrated in Figs.~\ref{fig:a2g-fault-vi}--\ref{fig:3ph-fault-vi}.
Each case started under the Unfaulted condition and then went into a faulted condition. After that the orchestrator was tasked with identifying the estimator model(s) that correspond to the different time stamps associated with each case, thereby predicting the operational conditions of the system.

\begin{table}[!htbp]
\centering
\setlength{\tabcolsep}{10pt} 
\renewcommand{\arraystretch}{1.25} 
\caption{Sequence of Events in Test Case}
\begin{tabular}{|c|c|c|c|} 
\hline
& \textbf{0 -- 0.25 sec} & \textbf{0.25 -- 0.5 sec} \\
\hline \hline
\textbf{Case I.} & Unfaulted & A-G fault\\ 
\textbf{Case II.} & Unfaulted & B-C fault   \\
\textbf{Case III.} & Unfaulted & 3-$\phi$ fault  \\
\hline
\end{tabular}
\label{table:Fault-cases}
\end{table}

\newpage
\section{Simulation Results} \label{sec:results}
\indent

\begin{figure}[!htbp]
\centering
\includegraphics[width=0.485\textwidth]{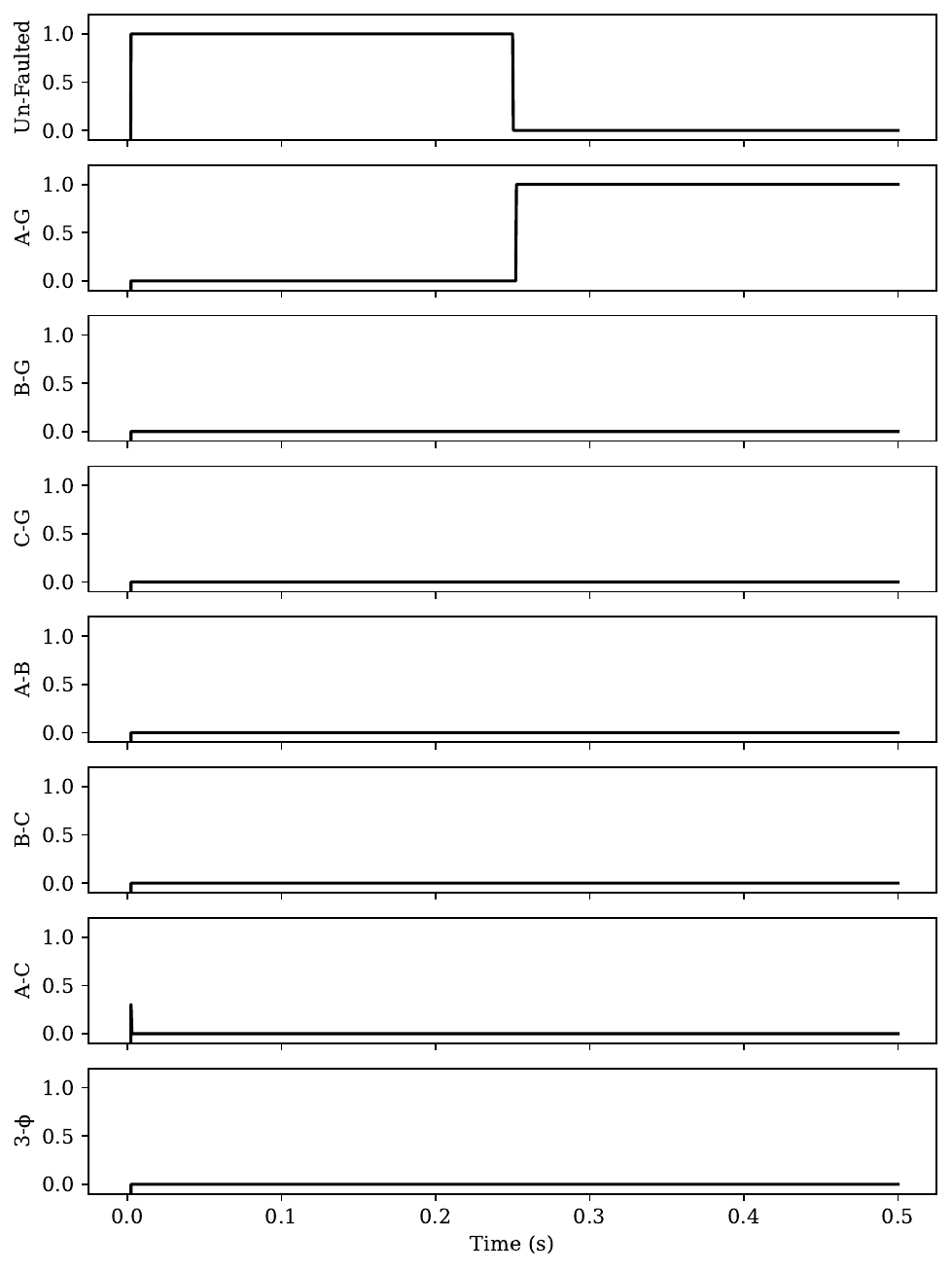}
\caption{Chi-squared confidence for all fault models during Case I.}
\label{fig:a2g-conf-all-models}
\end{figure}

\begin{figure}[!htbp]
\centering
\includegraphics[width=0.48\textwidth]{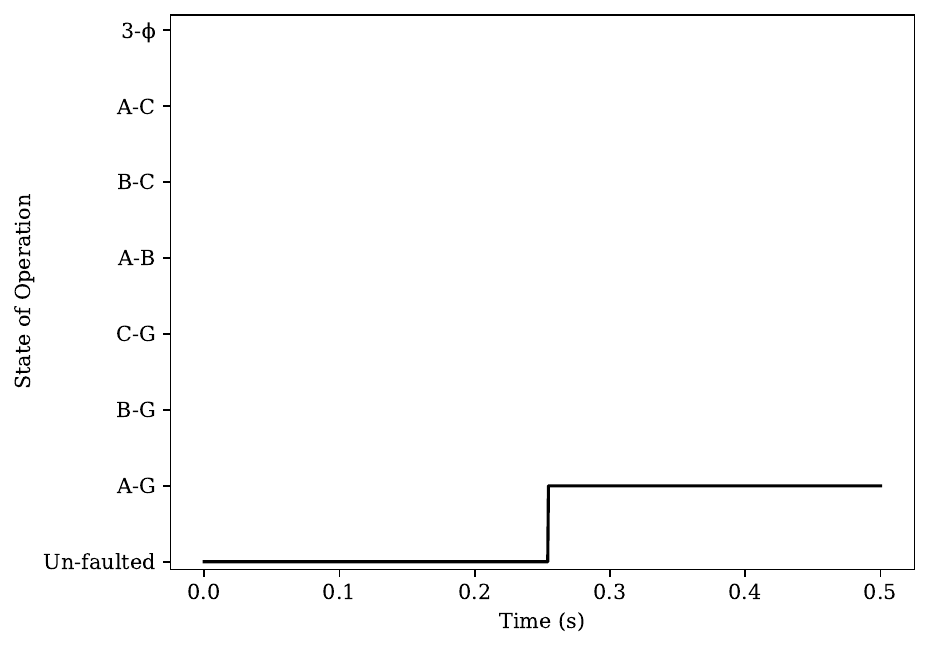}
 \caption{Fault models chosen by orchestrator during Case I.}
\label{fig:orch-a2g-fault}
\end{figure}

\begin{figure}[!htbp]
\centering
\includegraphics[width=0.485\textwidth]{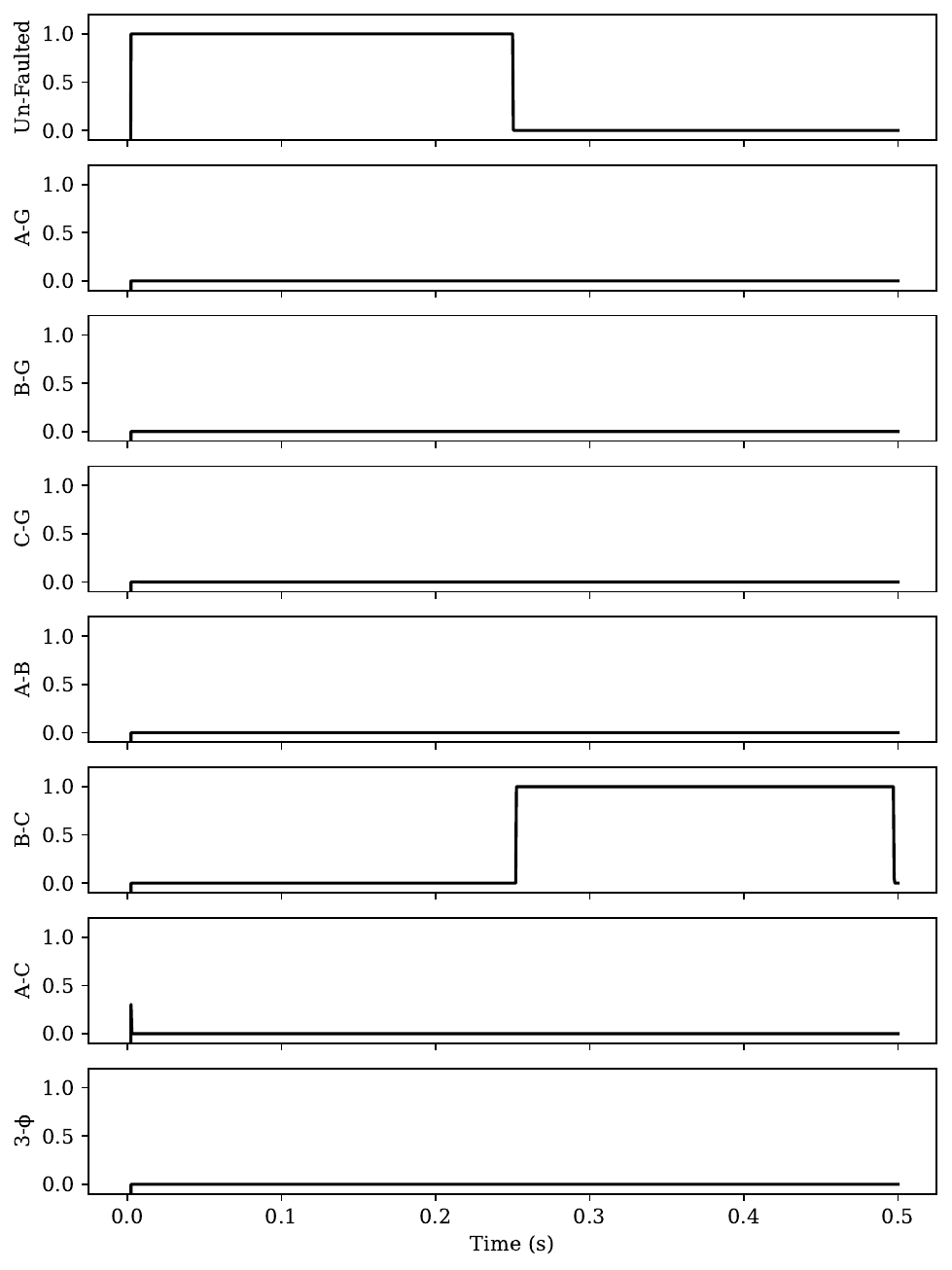}
\caption{Chi-squared confidence for all fault models during Case II.}
\label{fig:b2c-conf-all-models}
\end{figure}

\begin{figure}[!htbp]
\centering
\includegraphics[width=0.485\textwidth]{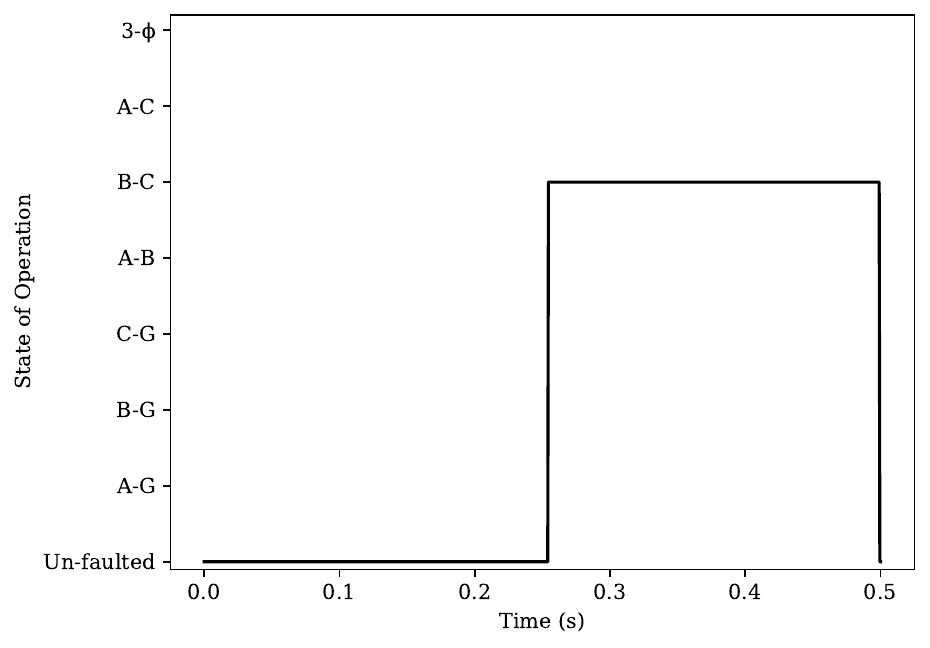}
\caption{Fault models chosen by orchestrator during Case II.}
\label{fig:orch-b2c-fault}
\end{figure}

\begin{figure}[!htbp]
\centering
\includegraphics[width=0.485\textwidth]{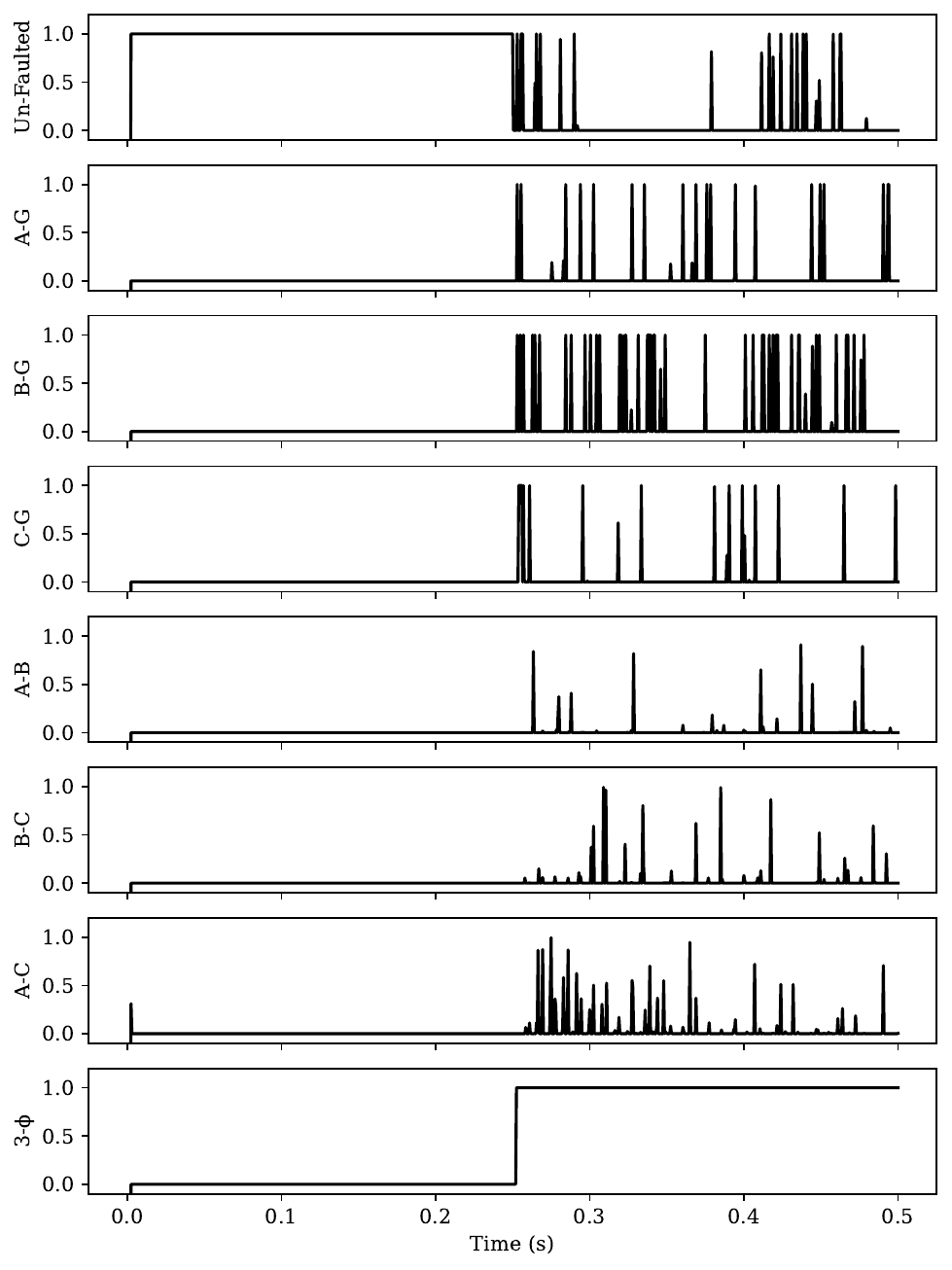}
\caption{Chi-squared confidence for all fault models during Case III.}
\label{fig:3ph-conf-all-models}
\end{figure}

\begin{figure}[!htbp]
\centering
\includegraphics[width=0.485\textwidth]{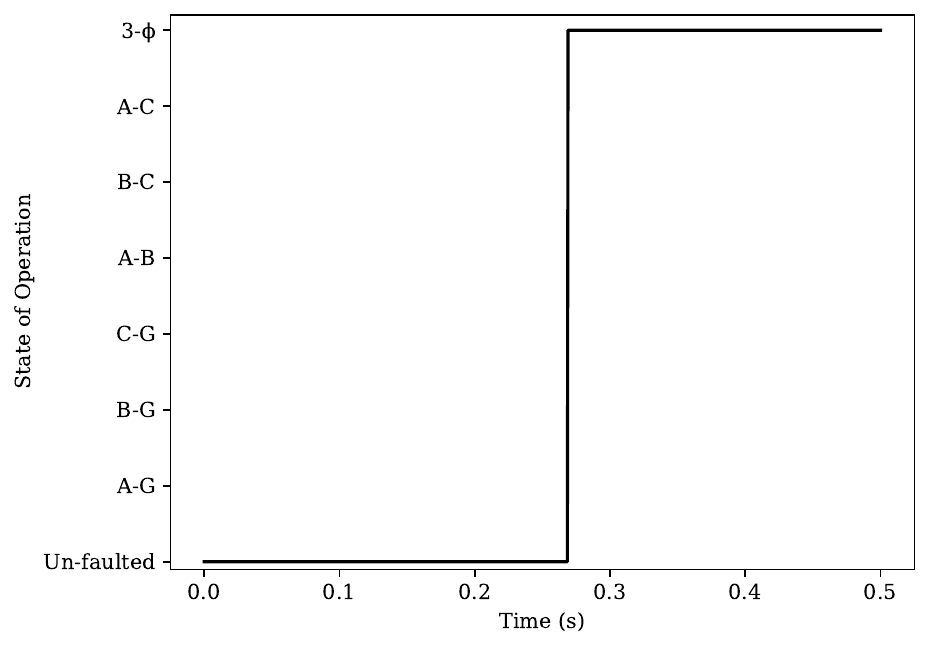}
\caption{Fault models chosen by orchestrator during Case III.}
\label{fig:orch-3ph-fault}
\end{figure}

Results verify the orchestrator's ability to be able to identify different operating conditions; illustrated on Fig.~\ref{fig:orch-a2g-fault}, Fig.~\ref{fig:orch-b2c-fault}, and Fig.~\ref{fig:orch-3ph-fault}.
The chi-squared statistics -- illustrated in Fig.~\ref{fig:a2g-conf-all-models}, Fig.~\ref{fig:b2c-conf-all-models}, and Fig.~\ref{fig:3ph-conf-all-models} -- were always considerably higher for the correct model than other models, thereby allowing easy distinction; therefore, the orchestrator was able to identify the right configuration and implement corresponding protection settings. 

\begin{figure}[!htbp]
\centering
\includegraphics[width=0.475\textwidth]{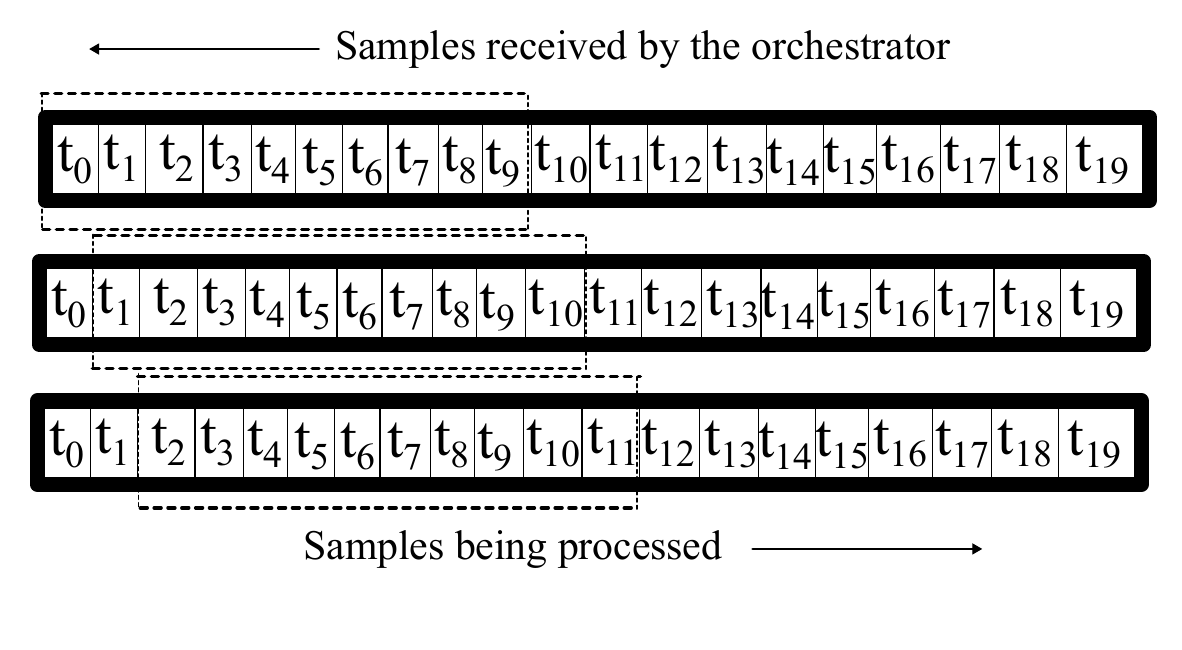}
\caption{Sliding window implementation.}
\label{fig:sliding-windown}
\end{figure}

\vspace{0.1in}
For the DSE implementation, a sliding-window approach was used -- as shown in Fig.~\ref{fig:sliding-windown} -- with 5 measurement samples considered for analysis to identify system status; the samples were collected every 500~$\mu $s, comparable to that of a PMU.
A hysteresis of 5 samples were used to offset erroneous model identifications during state transitions. 

\vspace{0.1in}
In Case~I and Case~II, the chi-squared confidences of relevant fault models were significantly higher than the others after the orchestrators processed measurements from post-fault conditions -- i.e., 2.5 ms (5 samples) post-fault -- which is comparably faster than a conventional relay trip scheme. The orchestrator, however, chose the correct fault model 9 samples later, owing to the hysteresis implementation.
There were no errors in the orchestrator's choice once all measurements corresponded to faulted conditions.

In Case~III, however, the implementation became complicated, since during the fault conditions several other models reported fairly high values of chi-squared confidence. This took a toll on the orchestator implementation and the model took significantly longer to identify the correct fault model; this was a delay of about 20ms (38 samples).
These spikes in chi-squared confidence for incorrect models during the faulted conditions were of short duration, and once the orchestrator identified the correct model, the hysteresis enabled these false positives to not have an impact.

Lastly, when the transition between Unfaulted and faulted conditions passed through the sliding window, there was a blackout period wherein the chi-squared confidences were small for all the models, thereby causing the orchestrator to not make a clear choice; the solution to this remains to be addressed.

\section{Conclusions} \label{sec:conclusions}
\indent

This paper investigated the implementation of a real-time application of DSE for the protection of inverter-interfaced microgrids.
The implementation, based on parameter estimation to determine system state, examined parallel processing of multiple models; DSE was able to identify the correct load configurations for measurements corresponding to normal and different modes of faulted operations.

The number of samples considered, along with the hysteresis, should also impact the choice of model and should potentially solve the experienced blackout issue. A delay in the availability of measurements, caused by channel delays, would impact the overall implementation; the orchestrator needs to be modified to address the same.

The impact of the ratio of the number of measurements to the number of states, and the influence of window size on estimation efficiency is yet to be addressed. To reduce model complexity, only balanced lumped radial loads were considered; further work is to be performed to investigate the impact of load imbalance on the quality of estimates. 

The time and computational complexity of running several models in parallel also require discussion, especially if measurements are received from a PMU and the inverter feeds a critical load. Finally, the impact of noise on the orchestrators performance has to be explored.


\bibliographystyle{unsrt}
\bibliography{references}

\end{document}